\documentclass[10pt, oneside]{article}   	
\usepackage{geometry}                		
\usepackage{graphicx}				
\usepackage{amssymb}
\usepackage[utf8]{inputenc} 
\usepackage[russian,english]{babel}
\usepackage{amsmath,hyperref}
\usepackage{amssymb}

\title{Defining Absence: The Origin of ``Neutrinoless''\\ and How it Obscures the Physics of Matter Creation}
\author{Francesco Vissani\\
\small INFN, Laboratori Nazionali del Gran Sasso}
\date{}

\begin{document}
\maketitle

\begin{abstract}
The term `neutrinoless’ is a cornerstone of modern particle physics, yet it defines a fundamental process by what is missing rather than what is created. We trace the origins of this privative neologism to a 1953 experimental claim and show how a `sociology of suspicion’ transformed Ettore Majorana’s affirmative ontology into an agnostic shorthand. By examining this linguistic shift, we argue that our current terminology may obscure the profound physical meaning of the search. Reclaiming the language of `matter creation’ is not merely a semantic choice, but a timely conceptual shift to bridge the gap between experimental caution and the radical character of the laws of nature we aim to uncover.

\end{abstract}

\vspace{0.0cm} 
\begin{center}
\itshape{\small In memory of Giorgio Giacomelli: a guide, an inspiration, a physicist.}
\end{center}
\vspace{0.2cm}

\subsection*{Introduction}

Experiments around the world are searching for an ultra-rare process dubbed {\em neutrinoless double beta decay.} These experiments are testing the theory of Ettore Majorana, who proposed that neutrinos coincide with their own antiparticles. This means that they can be reabsorbed in the nucleus, with only a pair of electrons being emitted. Testing this theory is important because it provides a unique chance to study matter creation in a laboratory setting, thereby pushing the boundaries of the Standard Model of particles and interactions. It may also provide clues about the origin of cosmic baryonic asymmetry.  However, the current terminology — especially the term `neutrinoless' — does not reflect the theoretical connection in any way and is not the original terminology. This essay aims to trace its historical origin.

The \textit{Oxford English Dictionary} (OED) currently dates the earliest evidence for the term \textit{neutrinoless} to 1969, citing an abstract in \textit{Physical Review Letters} \cite{prl}. However, this was not the terminology originally used by W. H. Furry in 1939 \cite{fu}
to describe the hypothetical process that has fascinated physicists for nearly a century. 
This brief note demonstrates that the term \textit{neutrinoless} actually dates back to the early 1950s---nearly two decades earlier than the OED record. This linguistic shift was not a mere shorthand, but a transition that reflects a profound change in the community's epistemological stance toward the nature of the neutrino.

\subsection*{From Majorana's theory to ``semantic bleaching'' (1939--1952)}

\begin{figure}[t] 
   \centering
   \includegraphics[width=\textwidth]{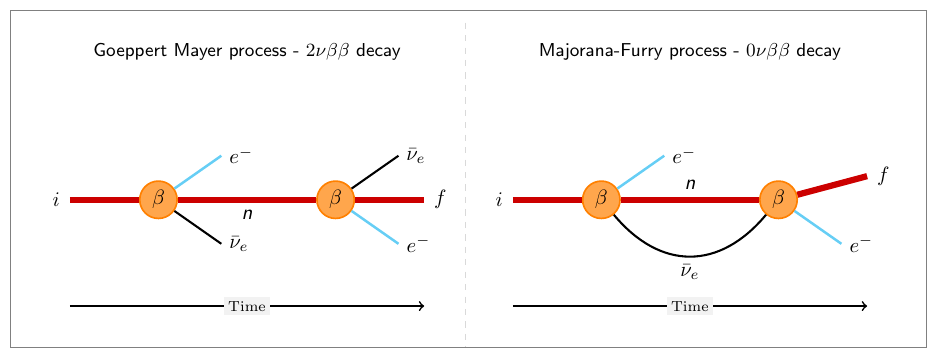} 
   \caption{\footnotesize \textbf{The two modes of double beta decay.}
(Left) The standard Goeppert-Mayer process  ($2\nu\beta\beta$), where two antineutrinos are emitted, preserving lepton number balance. (Right) The {Majorana-Furry process}  ($0\nu\beta\beta$), or ``neutrinoless'' decay. Both diagrams show the transition from an initial nucleus $i$ 
 to a final state $f$   via a virtual intermediate state $n$. While in the left panel antineutrinos carry away energy and compensate lepton number, in the right panel they are replaced by a virtual particle exchange between nucleons. This Majorana-Furry mode represents the net {creation of matter}, manifesting as the production of two electrons ($e^-$) without any compensating antimatter.}
   \label{fig1}
\end{figure}

In 1939, Furry introduced a new process characterized by the emission of two high-energy electrons (beta rays). He demonstrated that this channel was admissible under Ettore Majorana’s theory of the neutrino \cite{ma} and argued it would be significantly easier to observe than the channel with two neutrinos hypothesized by Maria Goeppert-Mayer \cite{gm} 
(see Figure~\ref{fig1} for a visual comparison of these two modes). 
During this initial phase, the language remained theoretically anchored: the community spoke explicitly of ``double beta decay in Majorana's theory.''

The connection between experimental signal and theoretical expectation began to fray in the early 1950s. The enthusiasm generated by certain optimistic expectations was initially reinforced by positive (and, in retrospect, spurious \cite{tr}) experimental results. In those years, the fundamental `bookkeeping’ of subatomic particles—the rules that dictate why matter doesn’t simply vanish or appear out of nowhere—was still being written. Indeed, the effort to formalize these very laws of conservation (what we call now `lepton number') was largely shaped by the attempt to categorize Furry's process \cite{ln1,ln2,ln3}.  This landscape of nascent laws and experimental ambiguity initiated what linguists call “semantic bleaching”—a process where a term loses its specific theoretical intensity.

A precursor to the modern neologism appeared in 1952, when H. Primakoff referred to the ``emission of two electrons without any accompanying neutrinos'' \cite{pr}. This descriptive phrase served as the semantic model for the more concise, but theoretically agnostic, term that was about to emerge.

\begin{figure}[t] 
   \centering
   \includegraphics[width=0.8\textwidth]{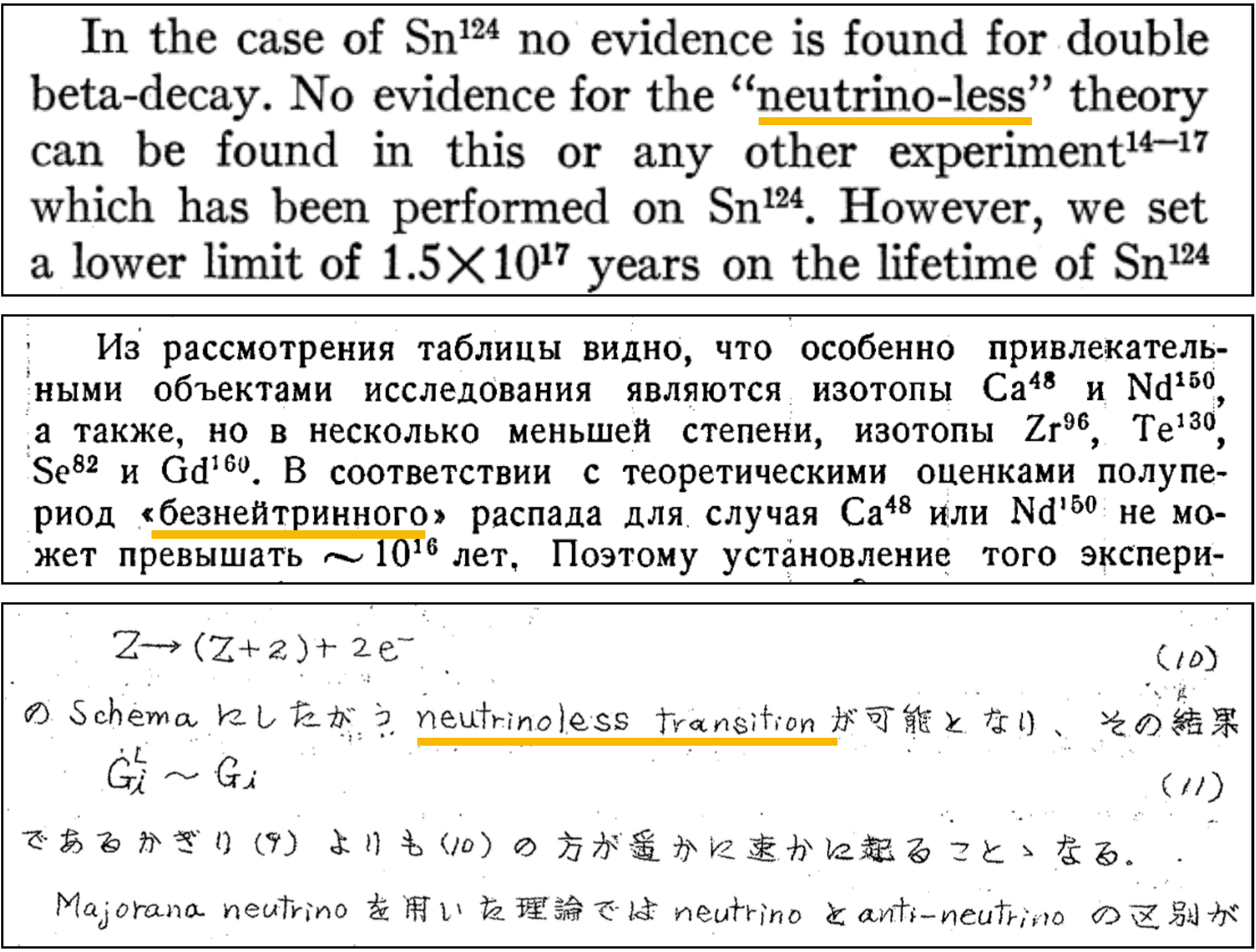} 
   \caption{\footnotesize \textbf{. The Rise of a Misnomer (1953-1955).} These brief excerpts from by the papers of McCarthy \cite{mc} (top panel), Zeldovich and collaborators \cite{ze} (middle panel), Sakata \cite{se} (bottom panel) show how, in a situation of great confusion (amidst partial experimental evidence and theories that were not yet complete) the term ``neutrino-less” quickly spread around the world and became established.}
   \label{fig2}
\end{figure}

\subsection*{The birth and global spread of the neologism (1953--1960)}

The term in use today first appeared in a 1953 experimental paper by J. A. McCarthy \cite{mc}. Notably, McCarthy’s work does not mention Majorana, Dirac, or Fermi; instead, it refers directly to evidence for the ``neutrino-less'' theory. In his conclusions, McCarthy states that his results ``indicate, but do not prove, that double beta-decay may occur in $^{96}$Zr without the emission of neutrinos.'' This phrasing reflects a shift toward a purely empirical description, stripping the process of its specific theoretical lineage.

The new term spread with remarkable speed. By 1954, it appeared in the doctoral thesis of chemist Henry Salig \cite{se}, who wrote: ``This suggests that neutrinoless double beta decay occurs.'' Simultaneously, the terminology crossed the Iron Curtain. A 1954 Russian review by Zeldovich and collaborators \cite{ze}
consistently employed the adjective \textit{bezneytrinnyy} 
(\foreignlanguage{russian}{безнейтринный})---a literal semantic twin of the English ``neutrinoless.''

A particularly striking attestation of this global shift is found in the 1955 work of Shoichi Sakata \cite{se}. Writing in Japanese, Sakata fiercely criticized what he perceived as the ``Americanization'' of science and expressed skepticism regarding the experimental verification of Majorana’s theory. Yet, in a linguistic incongruity, he used the English expression ``neutrinoless transition''---adopting the very terminology introduced by McCarthy to critique the field's direction
(the story is resumed in Figure~\ref{fig2}).

By 1960, the neologism had achieved total dominance. Influential reviews in \textit{Annals of Physics} \cite{gr} and \textit{Nuovo Cimento} \cite{fi} systematically adopted ``neutrinoless,” signaling that a term born from experimental caution had successfully eclipsed its ontological predecessor. Paradoxically, while the name flourished, the very experimental evidence that had propelled its rise—McCarthy’s 1953 excess—eventually proved spurious. The term remained as a linguistic monument to a vanished signal, further decoupling the language of the field from its theoretical foundations.

\begin{figure}[t!] 
   \centering
   \includegraphics[width=0.95\textwidth]{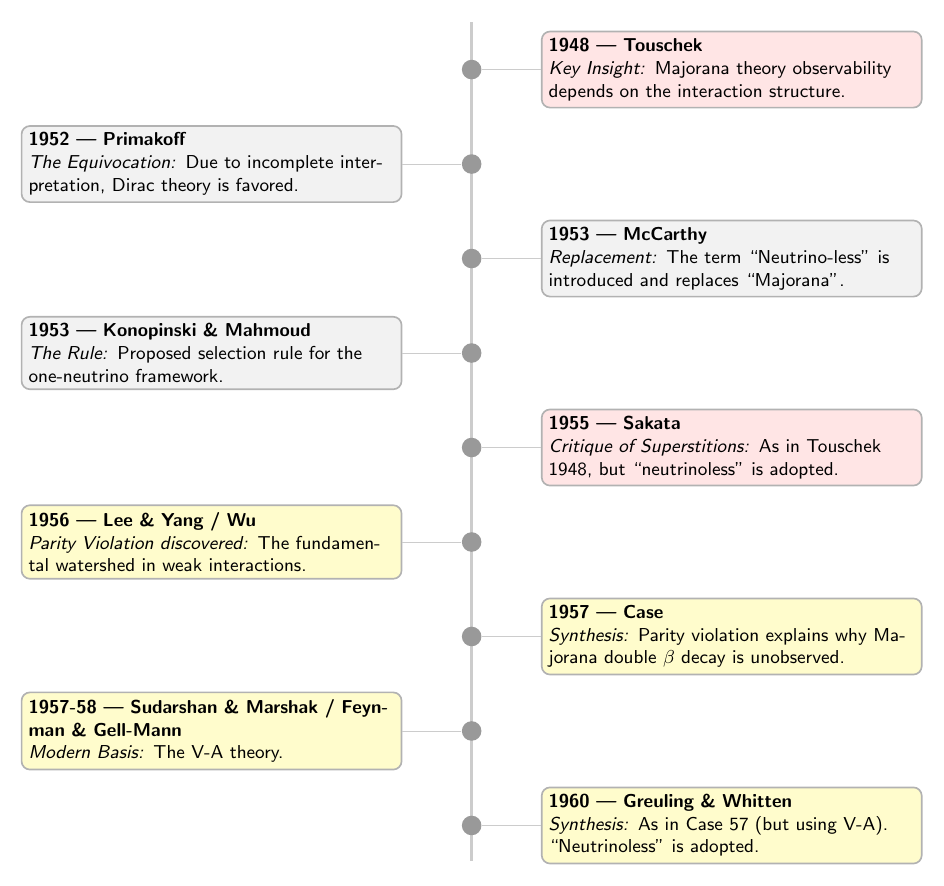} 
   \caption{\footnotesize  \textbf{Conceptual shifts in Majorana double beta decay (1948–1960).}
The timeline maps the historiographical fragmentation of the field through three thematic phases: 
\textit{Red boxes} (Touschek, Sakata): Theoretical anchors that reaffirmed Majorana’s vision, arguing that observability is intrinsically linked to the physical structure of the interaction.
\textit{Gray boxes} (Primakoff, McCarthy, Konopinski \& Mahmoud): The emergence of a {``phenomenological retreat,''} where the agnostic term ``neutrino-less'' began to eclipse the affirmative Majorana ontology in the experimental vocabulary. 
\textit{Yellow boxes} (Lee \& Yang, Wu, Case, Feynman \& Gell-Mann): The seismic impact of the {parity revolution} and the V-A theory. This era led to a dramatic redefinition of the experimental goal, effectively insulating the search for Majorana particles within a mathematical framework of suppression.}
   \label{fig3}
\end{figure}

This neologism is also a stigma of a period of profound conceptual fragmentation. Building upon the 1956 parity revolution \cite{ly}, Kenneth Case demonstrated that the decay is significantly slower than Furry had envisioned, being suppressed by the square of the Majorana neutrino mass \cite{ca}. This suppression defines the modern experimental challenge, yet his profound realization arrived in a climate of theoretical disarray and went initially unnoticed.  Struggling to reconcile new conservation laws with ambiguous signals, the community retreated into an agnostic label. ‘Neutrinoless’ became a safe harbor in the storm: by defining the process by what was missing, physicists could bypass the deepening mystery of its nature, effectively insulating the experimental hunt from the radical implications of the new theory 
(the timeline of this conceptual fragmentation is mapped in Figure~\ref{fig3}).

\subsection*{Beyond the sociology of suspicion: from absence to matter creation}

The linguistic divide between ``Majorana double beta decay'' and the ubiquitous ``neutrinoless'' variety reflects a profound shift in how we approach new physics. Linguistically, the suffix \textit{-less} provides a privative definition, identifying a phenomenon by what is absent. This label is a masterpiece of scientific empiricism: by focusing on the missing neutrino signal, it allows experimentalists to remain agnostic about the underlying mechanism.

However, this retreat reflects a recurring hope in physics: that data might eventually speak for themselves, free from the ``burden'' of a specific theoretical ontology. The shift toward ``neutrinoless'' in the 1950s was driven by what we might call a \textit{sociology of suspicion}. It was a collective retreat into safe, agnostic language---a linguistic firewall built to protect the community from the embarrassment of past false signals. But by choosing a word defined by what is \textit{not} there, we inadvertently severed the vital link to Majorana’s bold claim and the minimality of nature he envisioned in 1937. To treat the search merely as a hunt for an anomalous statistical peak, rather than the observation of matter creation, is to risk missing the very laws of nature we are trying to uncover.

We are now at a turning point. Recent discussions at the \textit{ECT* in Trento} \cite{ect} suggest that the community is ready to look beyond this historical artifact of caution. We are no longer just looking for a ``missing signal'' in a noisy background; we are explicitly discussing the experimental requirements to certify the birth of new matter in our detectors. We are after the physical appearance of two electrons in what would otherwise be a standard nuclear transformation of neutrons: a literal \textit{creation of matter} in the laboratory. Explicitly naming this process as such shifts the focus from an elusive absence to the observation of a net increase of matter particles in the universe \cite{vi,vi2}.

Conversely, fostering a detachment between theory and experiment is a perilous operation. Moving forward requires the courage to rely on our theoretical compass---from Majorana's original insight to the frontiers of particle physics. Without a reliable conceptual anchor, even a groundbreaking experimental result risks remaining a silent numerical anomaly. To bridge the gap in our understanding, we must reclaim a language that reflects the {radical nature of our inquiry}: moving from the study of an absence to the observation of one of the most fundamental events in the universe---the laboratory birth of new matter.
 
%

\subsection*{Acknowledgments}
{\footnotesize
With the partial support of the research grant 2022E2J4RK {\em PANTHEON: Perspectives in Astroparticle and Neutrino THEory with Old and New Messengers,} PRIN 2022 program funded by the Ministero dell'Università e della Ricerca'' (MUR). I thank
Angela Bracco, 
Osvaldo Civitarese, 
Luigi Coraggio, 
Stefano Dell'Oro, 
Eligio Lisi, 
Javier Men\'endez  
and 
Stefan Sch\"onert 
for valuable discussions.

}

\newpage
\small

\end{document}